\documentclass[preprint]{aastex}
\usepackage{comment}

\shorttitle{Multi-wavelength  Observations of Flaring Gamma-ray Blazar 3C 66A}
\shortauthors{Abdo et al.}

\begin{document}

%% one line. However, you may use \\ to force a line break ifs

\title{Multi-wavelength  Observations of the Flaring Gamma-ray Blazar \\ 3C 66A in 2008 October}


\author{
A.~A.~Abdo\altaffilmark{1,2}, 
M.~Ackermann\altaffilmark{3}, 
M.~Ajello\altaffilmark{3}, 
L.~Baldini\altaffilmark{4}, 
J.~Ballet\altaffilmark{5}, 
G.~Barbiellini\altaffilmark{6,7}, 
D.~Bastieri\altaffilmark{8,9}, 
K.~Bechtol\altaffilmark{3}, 
R.~Bellazzini\altaffilmark{4}, 
B.~Berenji\altaffilmark{3}
% ...
}

\altaffiltext{1}{Space Science Division, Naval Research Laboratory, Washington, DC 20375, USA}
\altaffiltext{2}{National Research Council Research Associate, National Academy of Sciences, Washington, DC 20001, USA}
\altaffiltext{3}{W. W. Hansen Experimental Physics Laboratory, Kavli Institute for Particle Astrophysics and Cosmology, Department of Physics and SLAC National Accelerator Laboratory, Stanford University, Stanford, CA 94305, USA}
\altaffiltext{4}{Istituto Nazionale di Fisica Nucleare, Sezione di Pisa, I-56127 Pisa, Italy}
\altaffiltext{5}{Laboratoire AIM, CEA-IRFU/CNRS/Universit\'e Paris Diderot, Service d'Astrophysique, CEA Saclay, 91191 Gif sur Yvette, France}
\altaffiltext{6}{Istituto Nazionale di Fisica Nucleare, Sezione di Trieste, I-34127 Trieste, Italy}
\altaffiltext{7}{Dipartimento di Fisica, Universit\`a di Trieste, I-34127 Trieste, Italy}
\altaffiltext{8}{Istituto Nazionale di Fisica Nucleare, Sezione di Padova, I-35131 Padova, Italy}
\altaffiltext{9}{Dipartimento di Fisica ``G. Galilei", Universit\`a di Padova, I-35131 Padova, Italy}
% ...

\newpage

\begin{abstract} The  BL Lacertae object  3C~66A was detected in a flaring state by the {\em Fermi} Large Area Telescope (LAT) and VERITAS  in   2008 October. In addition to these gamma-ray observations, F-GAMMA, GASP-WEBT, PAIRITEL, MDM, ATOM, {\em Swift}, and {\em Chandra} provided radio  to X-ray coverage. The available light curves show variability and, in particular, correlated flares are observed in the optical and {\em Fermi}-LAT gamma-ray band.  The resulting spectral energy distribution  can be well fit using standard leptonic models with and without an external radiation field for inverse-Compton scattering. It is found, however, that only the model with an external radiation field can accommodate  the intra-night variability observed at optical wavelengths.
 
 \end{abstract}



\section{Introduction}

The radio source 3C~66 \citep{3Ccatalog} was shown by \citet{Mackay1971} and \citet{Northover1973} to actually  consist of two unrelated radio sources separated by $0.11^\circ$ :  a compact source (3C~66A)  and a resolved galaxy (3C~66B).  3C~66A was subsequently identified as a quasi-stellar object by \citet{WillsandWills}, and as a BL Lacertae object   by  \citet{Smith1976}  based on its optical spectrum.  3C~66A is now a well-known blazar which, like other active galactic nuclei (AGN), is thought to be powered by  accretion of material onto a supermassive black hole located in the central region of the host galaxy \citep{UrryPadovani}. Some AGN present strong relativistic outflows in the form of jets, where particles are believed to be accelerated to ultra-relativistic energies and gamma rays are subsequently produced. Blazars are the particular subset of AGN with jets aligned to the observer's line of sight. Indeed,  the jet of 3C~66A has been imaged using very long baseline interferometry (VLBI) \citep{Taylor1996, Jorstad2001, Marscher2002, Britzen2007} and superluminal motion has been inferred \citep{Jorstad2001, Britzen2008}. This is indicative of the relativistic Lorentz factor of the jet and its small angle with respect to the line of sight.

%Due to its significant optical and X-ray variability, 3C~66A was identified as a BL Lacertae (BL Lac) object by  \citet{Maccagni87}. 

BL Lacs are known for having very weak (if any) detectable emission lines, which makes  determination of their redshift quite difficult. The redshift of 3C~66A was reported as $z=0.444$ by \citet{Miller78} and also  (although tentatively) by \citet{Kinney1991}. Each measurement however, is based on the measurement of a single line and is not reliable  \citep{bramel05}. Recent efforts (described in Section \ref{optical}) to  provide further constraints have proven unsuccessful.

Similar to other blazars, the spectral energy distribution (SED) of 3C~66A has two pronounced peaks, which suggests that at least two different physical emission processes are at work \citep{JoshiBoettcher}. The first peak, extending from radio to soft X-ray frequencies, is  likely due to  synchrotron emission from high-energy electrons, while different emission models  have been proposed to explain the second peak, which extends up to gamma-ray energies. Given the location of its synchrotron peak ($\lesssim10^{15}$ Hz), 3C~66A is further sub-classified as an intermediate synchrotron peaked blazar (ISP)  \citep{latSED}.

The models that have been proposed to explain gamma-ray emission in blazars can be roughly categorized into leptonic or hadronic, depending on whether the accelerated particles responsible for the gamma-ray emission are primarily electrons and positrons (hereafter ``electrons'') or protons. In leptonic models, high-energy electrons produce gamma rays via inverse Compton scattering of low-energy photons.  In synchrotron self-Compton (SSC) models, the same population of electrons responsible for the observed gamma rays generates the low-energy photon field through synchrotron emission. In external Compton (EC) models the low-energy photons originate outside the emission volume of the gamma rays. Possible sources  of target photons include: accretion-disk photons radiated directly into the jet \citep{DermerSchlickeiser}, accretion-disk photons scattered by emission-line clouds or dust into the jet \citep{SikoraBegelmanRees},  synchrotron radiation re-scattered back into the jet by broad-line emission clouds \citep{GhiselliniMadau}, jet emission from an outer slow jet sheet \citep{GhiselliniTavecchio2005}, or emission from faster or slower portions of the jet \citep{GeorganopoulosKazanas}. In hadronic models, gamma rays are produced by high-energy protons, either via proton synchrotron radiation \citep{Mucke}, or via secondary emission from photo-pion and photo-pair-production reactions (see \citet{Boettcher_review} and references therein for a review of blazar gamma-ray emission processes).

One of the main obstacles in the broadband study of gamma-ray blazars is the  lack of simultaneity, or 
at least contemporaneousness, of the data at the various wavelengths.  At high energies the situation is made even more difficult due to the lack of objects that can be detected by MeV/GeV and TeV observatories on comparable time scales. Indeed, until recently the knowledge of blazars at gamma-ray energies had been obtained from observations performed in two disjoint energy regimes:  i) the high energy (HE) range (20 MeV$<E<$ 10 GeV), studied in the 1990s by EGRET \citep{egret}, and ii) the very-high-energy (VHE) regime (E $>$ 100 GeV) observed by ground-based instruments \citep{Trevor_review}. Only\footnote{Markarian 501 was  marginally detected by EGRET only during a few months in 1996 \citep{Kataoka1999}.} Markarian 421 was  detected by both  EGRET and the  first   imaging atmospheric Cherenkov  telescopes  \citep{WhippleUpperLimits}. Furthermore, blazars detected by EGRET at MeV/GeV energies are predominantly flat-spectrum radio quasars (FSRQs), while  TeV blazars are, to date, predominantly  BL Lacs. It is important to understand these observational differences since they are likely related to the physics of the AGN \citep{cavaliere02}, or to the evolution of blazars over cosmic time \citep{BoettcherDermer}. 

The current generation of gamma-ray instruments (AGILE, {\em Fermi}, H.E.S.S., MAGIC and VERITAS) is closing the gap between the two energy regimes due to improved instrument sensitivities, leading us towards a deeper and more complete characterization of blazars as high-energy sources and as a population \citep{FermiTeV}.  An example of the successful synergy of space-borne and ground-based observatories  is provided by the joint observations of 3C~66A by {\em Fermi} and VERITAS during its strong flare of  2008 October.  The  flare was originally reported by VERITAS \citep{veritas_atel_3c66a, veritas_3c66a}, and soon after,  contemporaneous variability was also detected at optical to infrared wavelengths \citep{gasp_atel}, and in the {\em Fermi}-LAT energy band \citep{fermi_atel_3c66a}. Follow-up observations were obtained at radio, optical and X-ray wavelengths in order to measure the flux and spectral variability of the source across the electromagnetic spectrum and to obtain a quasi-simultaneous SED. This paper reports the results of this campaign, including the  broadband spectrum and a model interpretation of this  constraining SED.

\section{Observations and Data Analysis}

\subsection{VERITAS} 

The Very Energetic Radiation Imaging Telescope Array System (VERITAS) is an array of four 12m diameter  imaging atmospheric Cherenkov telescopes (IACTs) in southern Arizona, U.S.A. \citep{veritas_description}. 3C~66A was observed with VERITAS for 14 hours from 2007 September through  2008 January and for 46 hours between 2008 September and 2008 November. These observations (hereafter 2007-2008 data) add up to  $\sim$32.8 hours of live time after data quality selection. The data were analyzed following the procedure described in \citet{veritas_description}.  

As reported in \citet{veritas_3c66a}, the average spectrum measured by VERITAS is very soft, yielding a photon index $\Gamma$ of 4.1 $\pm0.4_{stat} \pm 0.6_{sys}$ when fitted to a power law $dN/dE \propto E^{-\Gamma}$. The average integral flux above 200 GeV  measured by VERITAS is $(1.3\pm0.1)\times10^{-11}$ cm$^{-2}$ s$^{-1}$, which  corresponds to 6\% of the Crab Nebula's flux above this threshold. In addition, a strong flare with night-by-night VHE-flux variability was detected in  October 2008.
%(top panel of Figure \ref{longterm}). 
For this analysis the VERITAS spectrum is  calculated for the short time interval  October 8 -- 10, 2008  (MJD 54747-54749;  hereafter {\em flare} period),  and for a longer period corresponding to the {\em dark run\footnote{Imaging atmospheric Cherenkov telescopes (IACTs) like VERITAS do not operate on nights with bright moonlight. The series of nights between consecutive bright-moonlight periods is usually referred as a {\em dark run}.}}  where most of the VHE emission from 3C~66A was detected (MJD 54734 - 54749). It should be noted that the {\em flare} and {\em dark run} intervals overlap and are therefore  not independent.  Table \ref{veritas} lists the relevant information from each data set.

 As shown in Figure \ref{gamma_sed}, the {\em flare} and {\em dark run} spectra are very soft, yielding nearly identical photon indices of  $4.1\pm0.6_{stat}\pm0.6_{sys}$, entirely consistent with that derived from the full 2007-2008 data set.  The integral flux above 200 GeV for the {\em flare} period is $(2.5\pm0.4)\times10^{-11}$ cm$^{-2}$ s$^{-1}$, while  the average flux for the {\em dark run} period is $(1.4\pm0.2)\times10^{-11}$ cm$^{-2}$ s$^{-1}$.  The extragalactic background light (EBL) de-absorbed spectral points for the {\em dark run} calculated using the  optical depth values of \citep{franceschini08} and assuming a nominal redshift of $z=0.444$, are also shown in Figure \ref{gamma_sed}. These points are well fit by a power law function with  $\Gamma=1.9\pm0.5$.

\begin{table}[p]
\begin{center}\footnotesize
\begin{tabular}{|c|c|c|c|c|c|c|}
\hline
Interval & Live Time [hr] & $N_{\rm On}$ & $N_{\rm Off}$ & Alpha & Excess & Significance [$\sigma$] \\
\hline
\hline
{\em flare} & 6.0 & 1531 & 7072 & 0.121 & 678.3 & 18.0 \\
{\em dark run} & 21.2 &  3888 & 20452 & 0.125 &  1331.5 & 22.2 \\
 2007-2008 & 28.1 & 7257 & 31201 & 0.175 & 1791 & 21.1 \\
\hline
\end{tabular}
\end{center}
\caption{Results from VERITAS observations of 3C~66A. Live time corresponds to the effective exposure time after accounting for data quality selection. $N_{\rm On}$ ($N_{\rm Off}$) corresponds to the number of on(off)-source events passing background-rejection cuts. Alpha is the normalization of off-source events and the excess is equal to $N_{\rm On}- \alpha  N_{\rm Off}$. The significance is expressed in number of standard deviations and is calculated according to equation (17) of \citet{LiMa}. See \citet{veritas_3c66a} for a complete description of the VERITAS analysis.} 
  \label{veritas}
\end{table}

\subsection{{\em Fermi}-LAT} 

The Large Area Telescope (LAT) on board the {\em Fermi} Gamma-ray Space Telescope  is a pair-conversion detector sensitive to gamma rays with energies between 20 MeV and several hundred GeV \citep{LAT}. Since launch  the instrument has  operated almost exclusively in sky survey mode, covering the whole sky every 3 hours. The overall coverage of the sky is fairly uniform, with exposure variations of $\le15\%$ around the mean value.  The LAT data  are analyzed  using ScienceTools v9r15p5 and instrument response functions P6V3 (available via the {\em Fermi} science support center\footnote{http://fermi.gsfc.nasa.gov/ssc/data/analysis/scitools/overview.html}). Only photons in the {\em diffuse} event class  are selected for this analysis because of their reduced charged-particle background contamination and very good angular reconstruction.  A zenith angle $<$105$^{\circ}$  cut in instrument coordinates is used  to avoid gamma rays from the Earth limb. The diffuse emission from the Galaxy is modeled using a spatial model ($\tt gll\_iem\_v02.fit$) which was refined with {\em Fermi}-LAT data taken during the first year of operation. The extragalactic diffuse and residual instrumental backgrounds are modeled as an isotropic component and are included in the fit\footnote{http://fermi.gsfc.nasa.gov/ssc/data/access/lat/BackgroundModels.html}.  The data are analyzed with an unbinned maximum likelihood technique \citep{MattoxLikelihood} using the likelihood analysis software developed by the LAT team.

\begin{figure}[p]
\begin{center}
\includegraphics[width=0.95\textwidth]{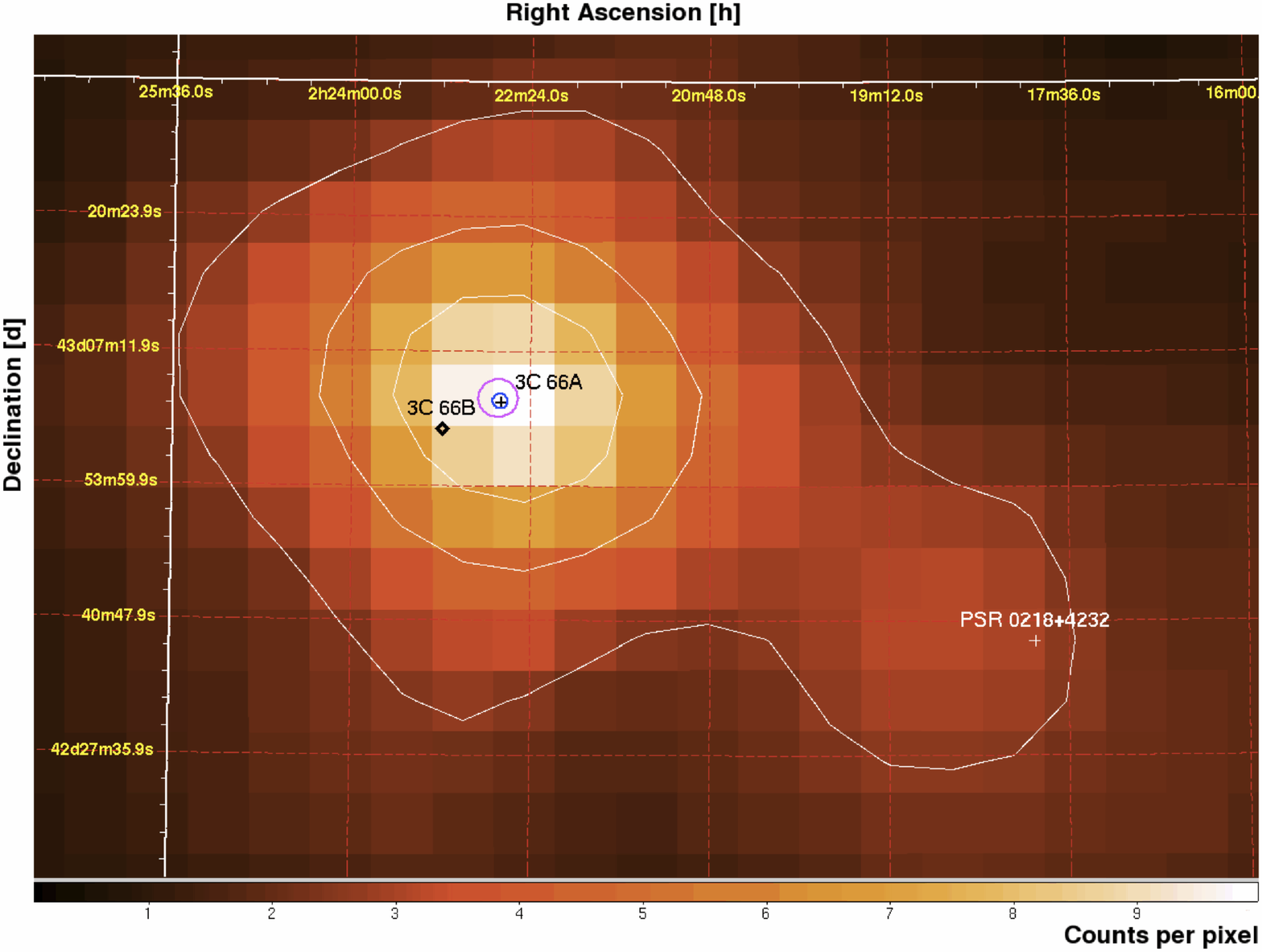}
\caption{Smoothed count map of the 3C~66A region as seen by {\em Fermi}-LAT between September 1, 2008 and December 31, 2008 with E$>$100 MeV. The color bar has units of counts per pixel and the pixel dimensions are  0.1$^\circ \times$ 0.1$^\circ$. The contour levels have been smoothed and correspond to 2.8, 5.2, and 7.6 counts per pixel.  The locations of 3C~66A and 3C~66B (a radio galaxy  that is $0.11^{\circ}$ away) are shown as a cross and as a diamond, respectively. The location of millisecond pulsar PSR 0218+4232 is also indicated with a white cross.   The magenta circle represents the VERITAS localization of the VHE source (RA;~DEC) = (2$^{\rm h}$~ 22$^{\rm m}$~41.6$^{\rm s}$~$\pm $1.7$^{\rm s}$$_{\rm stat}$~$\pm $6.0$^{\rm s}$$_{\rm sys}$ ; $43^{\rm o}$ 02'~ 35.5'' $\pm$ 21'' $_{\rm stat}$ $\pm$ 1'30'' $_{\rm sys}$) as reported in  \citet{veritas_3c66a}.  The blue interior circle represents the 95\% error radius of the {\em Fermi}-LAT localization (RA;~DEC) = (02$^{\rm h}$ 22$^{\rm m}$ 40.3$^{\rm s}$ $\pm$ 4.5$^{\rm s}$;  43$^{\rm o}$ 02' 18.6'' $\pm$ 42.1'') as reported in the {\em Fermi}-LAT first source catalog \citep{1FGL}. All positions are based on the J2000 epoch.}
\label{skymap}
\end{center}
\end{figure}

Although 3C~66A was detected by EGRET as source 3EG J0222+4253 \citep{3rdEgret}, detailed spatial and timing analyses by \citet{Kuiper00} showed that this EGRET source actually consists of the superposition of 3C~66A and the nearby millisecond pulsar PSR J0218+4232 which is  $0.96^{\circ}$ distant from the blazar. This interpretation of the EGRET data is verified by {\em Fermi}-LAT, whose improved angular resolution permits the clear separation of the two sources as shown in Figure \ref{skymap}. Furthermore, the known pulsar period is detected with high confidence  in the {\em Fermi}-LAT data \citep{millisecondPulsars}.  More importantly for this analysis, the clear separation between the pulsar and the blazar enables studies of each source independently in the maximum likelihood analysis, and thus permits an accurate determination of the spectrum and localization of each source, with negligible contamination.

Figure \ref{skymap} also shows the localization of the {\em Fermi} and VERITAS sources with respect to blazar 3C~66A and radio galaxy 3C~66B (see caption in Figure \ref{skymap} for details). It is clear from the map that the  {\em Fermi}-LAT and VERITAS localizations are  consistent and that the gamma-ray emission is confidently associated with the blazar and not with the radio galaxy. Some small contribution in the {\em Fermi}-LAT data from radio galaxy 3C~66B as suggested by  \citet{MAGIC_3c66b} and \citet{Tavecchio_Ghisellini} cannot be excluded  given the large spill-over of low-energy photons from 3C~66A at the location of 3C~66B. This is due to the long tails of the {\em Fermi}-LAT  point-spread-function at low energies as described in \citet{LAT}. Nevertheless,  considering only photons with energy $E>1$ GeV, the upper limit ($95\%$ confidence level) for a source at the location of 3C~66B is 2.9 x 10$^{-8}$cm$^{-2}$s$^{-1}$  for the {\em dark run} period (with a test statistic\footnote{The test statistic value $(TS)$ quantifies the probability of having a point  source at the location specified. It is roughly the square of the significance value: a TS of 25  corresponds to a signal of approximately 5 standard deviations \citep{1FGL}.} $TS = 1.3$).  For the 11 months of data corresponding to the first {\em Fermi}-LAT catalog \citep{1FGL}  the upper limit is  4.9 x 10$^{-9}$cm$^{-2}$s$^{-1}$  ($TS = 5.8$).

\begin{figure}[p]
\begin{center}
\includegraphics[width=0.8\textwidth]{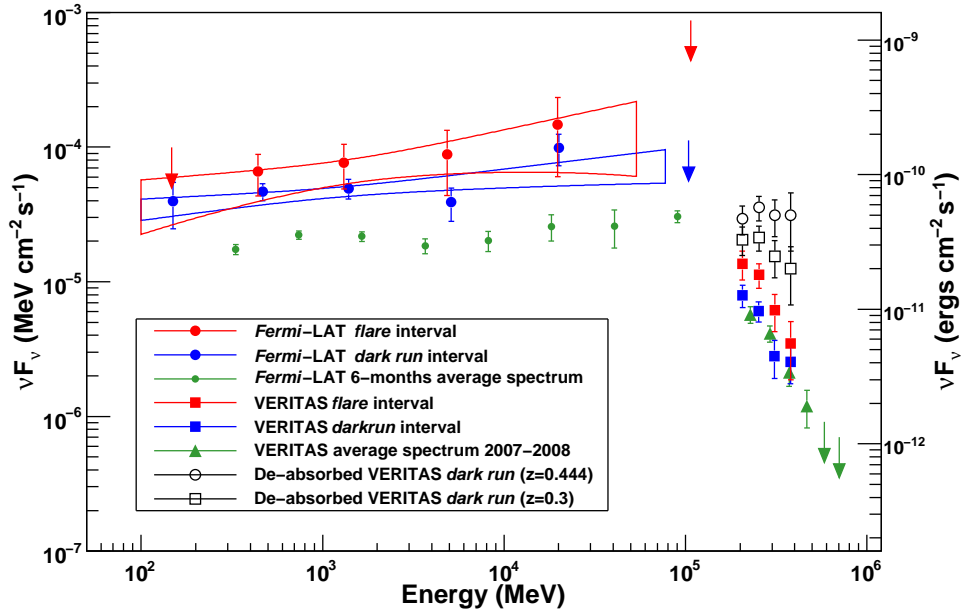}
\caption{{\footnotesize  Gamma-ray SED of 3C~66A including {\em Fermi}-LAT and VERITAS data for the {\em flare} (red symbols) and {\em dark run} (blue symbols) intervals. The {\em Fermi}-LAT spectra is also shown here as ``butterfly'' contours (solid lines) describing the statistical error on the spectrum \citep{FermiTeV}. The previously reported  {\em Fermi}-LAT 6-month-average spectrum \citep{LBASspectra} is also shown here (green circles) and is lower than the spectrum obtained during the campaign.  The average 2007--2008  VERITAS spectrum  originally reported in \citet{veritas_3c66a} is displayed with green triangles. In all cases the upper limits are calculated at 95\% confidence level. The de-absorbed {\em dark run} spectra obtained using the optical depth values of \citet{franceschini08} are also shown as open circles and open squares for redshifts of 0.444 and 0.3, respectively.}}
\label{gamma_sed}
\end{center}
\end{figure}

As in the analysis of  the VERITAS observations, the {\em Fermi}-LAT spectrum is calculated for the {\em flare}   and for the {\em dark run} periods. The {\em Fermi} {\em flare} period flux   $F(E>$100MeV) = $(5.0\pm1.4_{\rm stat}\pm0.3_{\rm sys})$ x 10$^{-7}$cm$^{-2}$s$^{-1}$  is consistent within errors with the {\em dark run} flux of $(3.9\pm0.5_{\rm stat}\pm0.3_{\rm sys})$ x 10$^{-7}$cm$^{-2}$s$^{-1}$. In both cases the {\em Fermi}-LAT spectrum is quite hard and can be described by a power law with a photon index $\Gamma$ of $1.8\pm0.1_{\rm stat}\pm{0.1}_{\rm sys}$ and  $1.9\pm0.1_{\rm stat}\pm{0.1}_{\rm sys}$  in the {\em flare} period and {\em dark run} intervals, respectively.   Both spectra are shown in the high-energy SED in Figure \ref{gamma_sed}. 

\subsection{{\em Chandra}}
3C~66A was observed by the {\em Chandra} observatory on October 6, 2008 for a
total of 37.6~ksec with the Advanced CCD Imaging Spectrometer (ACIS), covering
the energy band between 0.3 and 10~keV. The source was observed in the continuous clocking (CC) mode to avoid pile-up effects. Standard analysis tools (CIAO 4.1) and  calibration files (CALDB v3.5.0) provided by the Chandra X-ray center\footnote{http://cxc.harvard.edu/ciao/}  are used. 

The time-averaged spectrum is obtained and re-binned to ensure
 that each spectral channel contains at least 25 background-subtracted
 counts. This condition allows the use of the $\chi^2$ quality-of-fit
 estimator to find the best fit model.  XSPEC v12.4 \citep{Arnaud1996} is used  
 for the spectral analysis and fitting procedure. 

Two spectral models have been used to fit the data: single power law  and broken power law. Each model includes  galactic  H\,{\sc i} column density ( N$_{H, \rm Gal}=8.99\times10^{20}$~cm$^{-2}$) according to  \citet{Dickey90}, where the photoelectric absorption is set with the XSPEC model  {\em phabs}\footnote{http://heasarc.gsfc.nasa.gov/docs/software/lheasoft/xanadu/xspec/manual/XSmodelPhabs.html}. An additional local H\,{\sc i} column density was also tried but in both cases the spectra was consistent with pure galactic density. Consequently, the column density has been fixed to the galactic value  in each model and  the  obtained results are presented in Table \ref{chandra}. An F-test was performed to demonstrate that the spectral fit improves significantly when using the extra degrees of freedom of the broken power law model. Table \ref{chandra} also contains the results of the F-test.

\begin{table}[p]
\begin{center}\footnotesize
\begin{tabular}{|c|c|c|c|c|c|}
\hline
\hline
\multicolumn{6}{|c|}{Single Power Law Model}\\
\hline \hline
\multicolumn{2}{|c|}{ {\scriptsize  $\Gamma$}} &  \multicolumn{2}{|c|}{ {\scriptsize Flux [$10^{-12}$ ergs cm$^{-2}$ s$^{-1}$] }}&  \multicolumn{2}{|c|}{ {\scriptsize $\chi^2/d.o.f.$}} \\

\hline
\multicolumn{2}{|c|}{2.99$\pm$0.03}
& \multicolumn{2}{|c|}{3.47$\pm0.06$}
& \multicolumn{2}{|c|}{1.21 (232.6/193)}

\\
\hline
\hline
\multicolumn{6}{|c|}{Broken Power Law Model}\\
\hline
\hline
{\scriptsize  $\Gamma_1$} & {\scriptsize $\Gamma_2$}& {\scriptsize Flux [$10^{-12}$ ergs cm$^{-2}$ s$^{-1}$]}& {\scriptsize Break [keV]} & {\scriptsize $\chi^2/d.o.f.$} & {\scriptsize F-test Prob.}\\
\hline
 3.08$^{+0.3}_{-0.5}$
& 2.24$^{+0.23}_{0.37}$
& 3.58$^{+0.07}_{-0.08}$
&3.3$^{+0.5}_{-0.3}$
& 0.97 {(185.2/191)}
& 3.47 $\times$ 10$^{-10}$
\\

\hline
\end{tabular}
\end{center}
\caption{Best-fit model parameters for a fit performed to the {\em Chandra} data in the $1-7$ keV energy range. The galactic N$_{H,Gal}$ value is fixed to $8.99\times10^{20}$ cm$^{-2}$, the value of the galactic H\,{\sc i} column density according to \citet{Dickey90}.  Errors indicate the 90\% confidence level.}
  \label{chandra}
\end{table}

\subsection{{\em Swift} XRT and UVOT}  
Following the VERITAS detection of VHE emission from 3C~66A, Target of Opportunity (ToO) observations of 3C~66A with {\em Swift} were obtained for a total duration of $\sim$10 ksec. The {\em Swift} satellite observatory comprises an UV-Optical telescope (UVOT), an X-ray telescope (XRT) and a Burst Alert Telescope \citep{swift}. Data reduction and calibration of the XRT data are performed with {\em HEASoft} v6.5 standard tools. All XRT data presented here are taken in Photon Counting (PC) mode with negligible pile-up effects. The X-ray spectrum of each observation is fit with an absorbed power law using a fixed Galactic column density from \citet{Dickey90}, which gives good $\chi^{2}$ values for all observations. The measured photon spectral index ranges between 2.5 and 2.9 with a typical statistical uncertainty of 0.1.

UVOT obtained data through each of six color filters, {\em V}, {\em B} and {\em U} together with filters defining three ultraviolet passbands, {\em UVW1}, {\em UVM2} and {\em UVW2} with central wavelengths of 260 nm, 220 nm and 193 nm respectively. The data are calibrated using standard techniques \citep{Poole08} and  corrected for Galactic extinction by interpolating the absorption values from  \citet{Schlegel1998}  ($E_{B-V} = 0.083$ $ \rm mag$) with the galactic spectral extinction model  of \citet{Fitzpatrick1999}.

\subsection{\label{optical}Optical to Infrared Observations} 

The {\em R} magnitude  of the host galaxy of 3C~66A is $\sim$19   in the optical band \citep{wurtz1996}. Its contribution is negligible compared to the typical AGN magnitude of {\em R} $\lesssim$15, therefore  host-galaxy correction is not necessary.

{\bf GASP-WEBT:}  3C~66A is continuously monitored by telescopes affiliated to  the  GLAST-AGILE support program of the  Whole Earth Blazar Telescope (GASP-WEBT; see \citet{gasp, gasp2009}). These observations provide a long-term light curve of this object with complete sampling as shown in Figure \ref{longterm}. During the time interval in consideration (MJD 54700 - 54840), several observatories (Abastumani, Crimean, L'Ampolla, Lulin, New Mexico Skies,  Roque de los Muchachos (KVA), Rozhen, Sabadell, San Pedro Martir, St. Petersburg, Talmassons, Teide (BRT), and Tuorla) contributed photometric observations in the {\em R} band. Data in the {\em J}, {\em H} and {\em K} band were taken at the Campo Imperatore observatory.  A list of the observatories and their location is available in Table \ref{observatories}.

\begin{table}
\begin{center}\footnotesize
\begin{tabular}{|c c c|}
\hline
Observatory & Location & Web page \\
\hline
\hline
\multicolumn{3}{|c|}{Radio Observatories}\\
Crimean Radio Obs. & Ukraine & www.crao.crimea.ua \\
Effelsberg & Germany & www.mpifr.de/english/radiotelescope \\
IRAM & Spain & www.iram-institute.org/EN/30-meter-telescope.php \\
Medicina & Italy & www.med.ira.inaf.it \\
Mets\"ahovi & Finland & www.metsahovi.fi/en \\
Noto & Italy & www.noto.ira.inaf.it \\
UMRAO & Michigan, USA & www.astro.lsa.umich.edu/obs/radiotel \\
\hline
\multicolumn{3}{|c|}{Infrared Observatories}\\
Campo Imperatore & Italy & www.oa-teramo.inaf.it \\
PAIRITEL & Arizona, USA & www.pairitel.org \\
\hline
\multicolumn{3}{|c|}{Optical Observatories}\\
Abastumani & Georgia & www.genao.org \\
Armenzano & Italy & www.webalice.it/dcarosati \\
ATOM & Namibia & http://www.lsw.uni-heidelberg.de/projects/hess/ATOM/ \\
%Belogradchick & Bulgaria & http://www.astro.bas.bg/aobel/
%Obs. de Bourdeaux & France & http://www.obs.u-bordeaux1.fr/
%Catania & Italy & http://www.ct.astro.it/
Crimean Astr. Obs. & Ukraine & www.crao.crimea.ua \\
%Gualba Obs. &  Spain & 
%Jakokoski Obs. & Finland & http://cc.joensuu.fi/seulaset/ccd/observatory/
Kitt Peak (MDM) & Arizona, USA & www.astro.lsa.umich.edu/obs/mdm \\
L'Ampolla & Spain & \\
Lulin & Taiwan & www.lulin.ncu.edu.tw/english \\
%Michael Adrian Obs. & Germany & http://www.t1t-trebur.de/
New Mexico Skies Obs. & New Mexico, USA & www.nmskies.com \\
%Perugia & Italy & http://astro.fisica.unipg.it/
Roque (KVA) & Canary Islands, Spain & www.otri.iac.es/eno/nt.htm \\
Rozhen & Bulgaria &  www.astro.bas.bg/rozhen.html \\
Sabadell & Spain & www.astrosabadell.org/html/es/observatoriosab.htm \\
San Pedro M\'{a}rtir & M\'exico & www.astrossp.unam.mx/indexspm.html \\
St. Petersburg & Russia & www.gao.spb.ru \\
Talmassons & Italy & www.castfvg.it \\
Teide (BRT)  & Canary Islands, Spain & www.telescope.org \\
Torino & Italy & www.to.astro.it \\
Tuorla & Finland & www.astro.utu.fi \\
Valle d' Aosta & Italy & www.oavda.it/english/osservatorio \\
%Xinglong & China & 
\hline
\hline
\multicolumn{3}{|c|}{Gamma-ray Observatories}\\
VERITAS & Arizona, USA & veritas.sao.arizona.edu \\
\hline
\end{tabular}
\end{center}
\caption{List of ground-based observatories that participated in this campaign.}
  \label{observatories}
\end{table}

{\bf MDM:}  Following the discovery of VHE emission, 3C~66A was observed with the 1.3m telescope of the 
MDM Observatory during the nights of Oct. 6 - 10,
2008. A total of 290 science frames in {\em U}, {\em B},
{\em V},  and {\em R} bands (58 each) were taken throughout
the entire visibility period (approx. 4:30 -- 10:00
UT) during each night. The light curves, which cover the time around the flare, are presented in Figure \ref{shortterm}.

{\bf ATOM:}  Optical observations for this campaign in the R band were also obtained with  the 0.8 m optical telescope
ATOM in Namibia which monitors this source periodically.  Twenty photometric observations are available starting on MJD 54740 and are shown in Figures \ref{longterm} and \ref{shortterm}.
%Data analysis  is conducted automatically by pipeline processing using SExtractor \citep{SEextractor}. The
%magnitudes were converted in fluxes using the zeropoints of \citet{bessell1979}.  

{\bf PAIRITEL:}  Near-infrared observations in the  {\em J}, {\em H} and {\em $K_{s}$} were obtained following the VHE flare with the 1.3m Peters Automated Infrared Imaging Telescope (PAIRITEL; see  \citet{pairitel}) located at the Fred Lawrence Whipple Observatory.  The resulting light curves using differential photometry with four nearby calibration stars are shown in Figure \ref{shortterm}.

{\bf Keck:}  The optical spectrum of 3C~66A was measured with the LRIS spectrometer \citep{Oke} on the Keck~I telescope on the night of  2009 September 17 under good conditions.  
%The instrument was configured with the 600/7500 grating, the D560 dichroic, the B600/4000 grism, and a $1''$ longslit.  
The instrument configuration resulted in a full-width-half-maximum of $\sim 250$  km s$^{-1}$ over the wavelength range 3200--5500\AA\  (blue  side) and $\sim 200$ km s$^{-1}$ over the range 6350--9000\AA\ (red side). A series of exposures totaling 110  seconds (blue) and 50 seconds (red) were obtained,  yielding a signal-to-noise (S/N) per resolution element of $\sim 250$ and 230 for the blue and red cameras respectively. The data were reduced with the LowRedux\footnote{http://www.ucolick.org/$\sim$xavier/LowRedux/index.html} pipeline  and
calibrated using a spectrophotometric star observed on the same night. Inspection of the 3C~66A spectrum reveals no spectral features aside from those imposed by Earth's atmosphere and
the Milky Way (Ca H+K).  Therefore, these new data do not offer any insight on the redshift of 3C~66A and in particular are unable to confirm the previously reported value of $z=0.444$ \citep{Miller78}.

\subsection{Radio Observations}
Radio observations are available thanks to the F-GAMMA (Fermi-Gamma-ray Space Telescope AGN Multi-frequency Monitoring Alliance) program, which is dedicated to monthly monitoring of selected {\it Fermi}-LAT blazars \citep{Fuhrmann07, fgamma}. Radio flux density measurements were conducted with the 100-m Effelsberg radio telescope  at
4.85, 8.35, 10.45, and 14.60 GHz on 2008 October 16.  These data are supplemented with an additional measurement at 86 GHz conducted with
the IRAM 30-m telescope (Pico Veleta, Spain), on  2008 October 8. The data were reduced using standard procedures described in \citet{Fuhrmann08}.  Additional radio observations taken between October 5 and October 15, 2008  (contemporaneous to the {\em flare} period) are provided by the Medicina, Mets\"ahovi, Noto, and UMRAO observatories, all of which are members of the GASP-WEBT consortium.

\section{Discussion}

\subsection{Light Curves}

\begin{figure}[p]
\begin{center}
\includegraphics[width=0.8\textwidth]{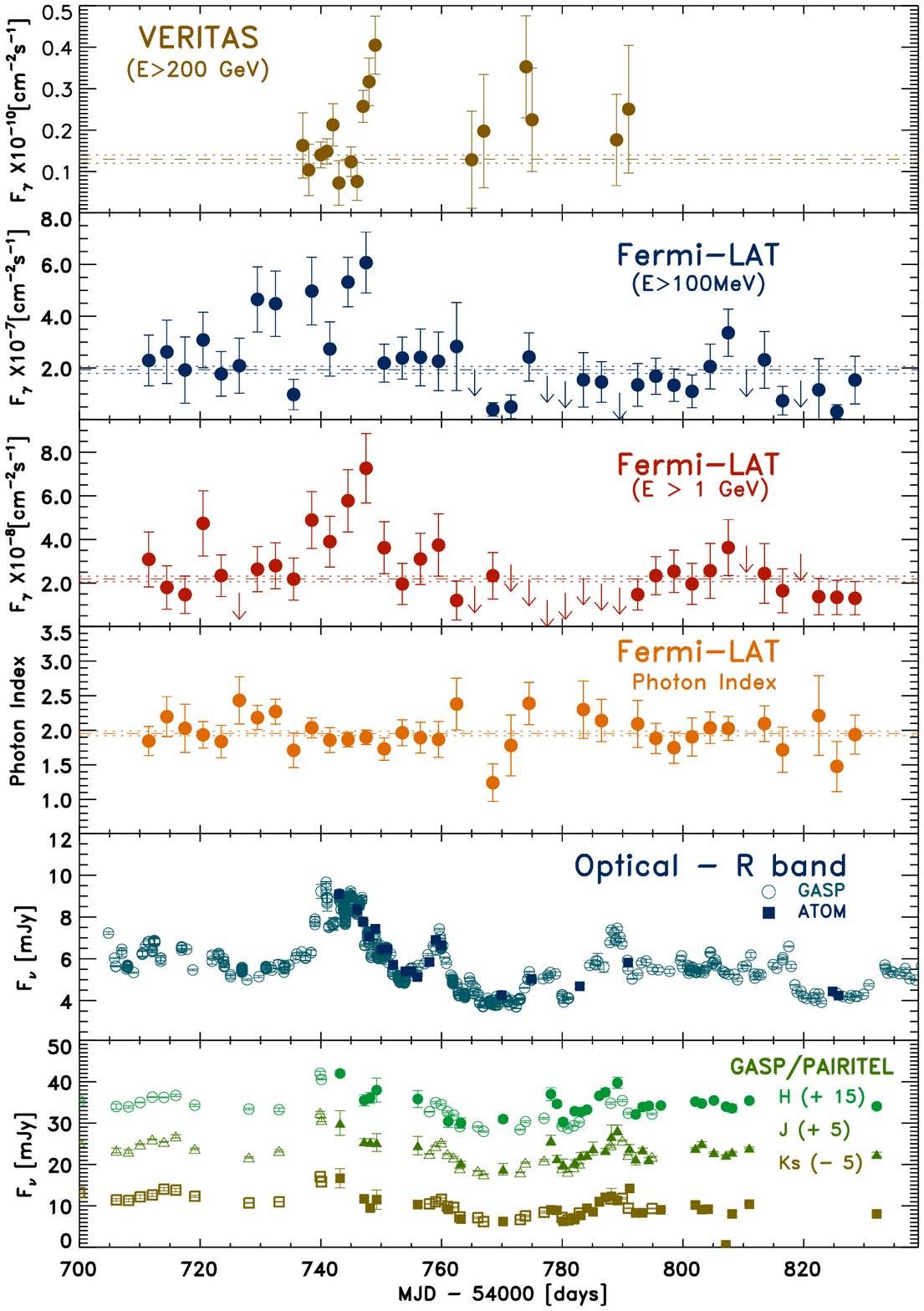}
\caption{{\footnotesize 3C~66A light curves covering  2008 Aug 22 to 2008 Dec 31 in order of increasing wavelength.  The VERITAS observations are combined to obtain nightly flux values and the dashed and dotted lines  represent the average flux measured from the 2007--2008 data and its standard deviation. The {\em Fermi}-LAT light curves contain  time bins with a width of 3 days. The average flux and average photon index measured by Fermi-LAT during the first six months of science operations are shown as horizontal lines in the respective panels. In all cases the {\em Fermi}-LAT photon index is calculated over the 100 MeV to 200 GeV energy range. The long-term light curves at optical and infrared wavelengths are presented in the two bottom panels. In the bottom panel GASP-WEBT  and PAIRITEL observations are represented by open and solid symbols, respectively.}}
\label{longterm}
\end{center}
\end{figure}

\begin{figure}[p]
\begin{center}
\includegraphics[width=0.8\textwidth]{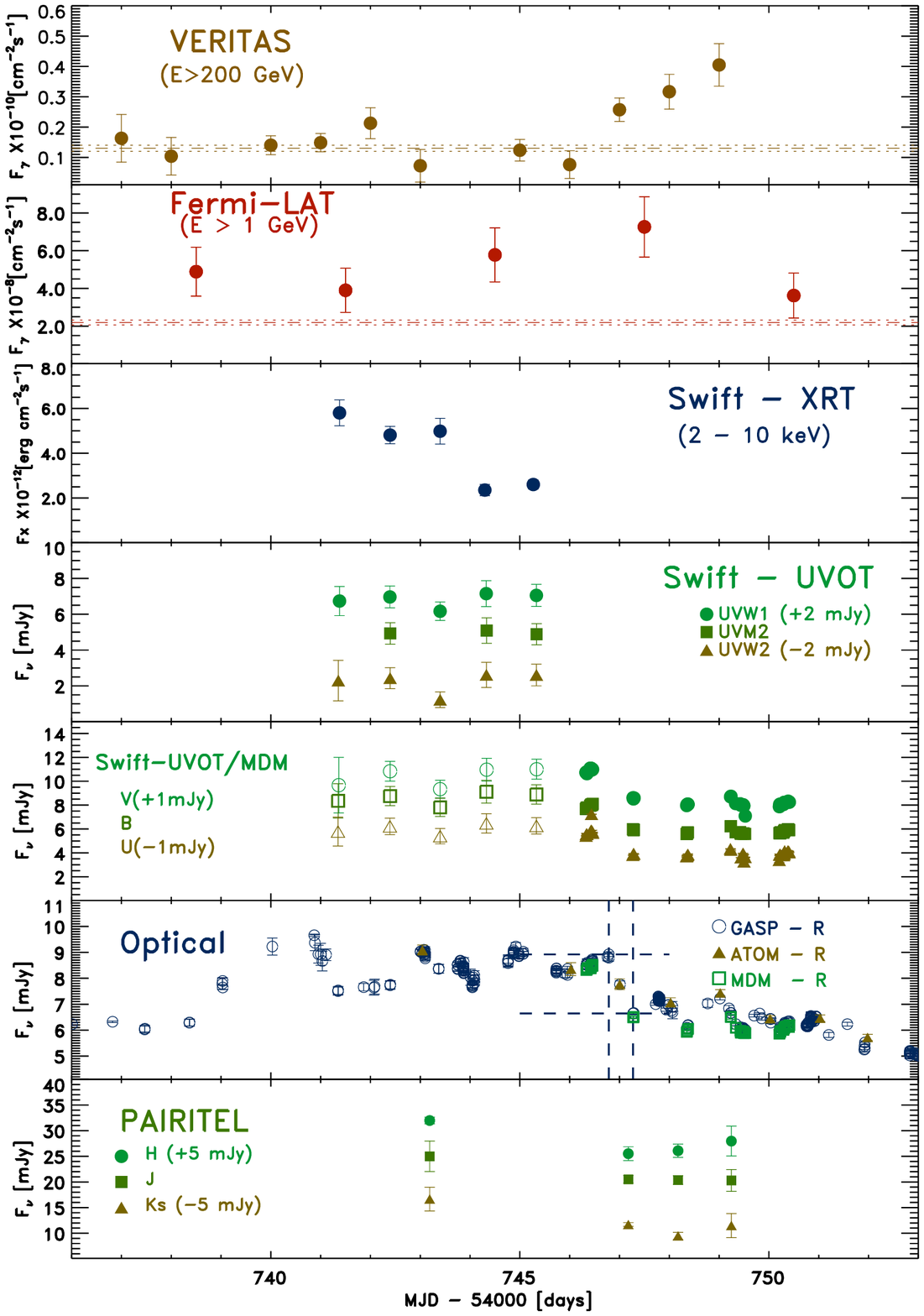}
\caption{{\footnotesize 3C~66A light curves covering the period centered on the gamma-ray flare (2008 Oct 1 - 10). The VERITAS and {\em Fermi}-LAT panels were already described in the caption of Figure \ref{longterm}. {\em Swift} Target-of-Opportunity (ToO) observations (panels 3-5 from the top) were obtained following the discovery of  VHE emission by VERITAS \citep{veritas_atel_3c66a}.  {\em Swift}-UVOT and MDM observations are represented by open and solid symbols, respectively.  The optical light curve in panel 6 from the top displays intra-night variability.  An example is identified in the plot, when  a rapid decline of the optical flux by $\Delta F/\Delta t \sim-0.2$ mJy hr$^{-1}$  is observed on MJD 54747.}}
\label{shortterm}
\end{center}
\end{figure}

The resulting multiwavelength light curves from this campaign are shown in Figure \ref{longterm} for those bands with long-term coverage and in Figure \ref{shortterm} for those observations that were obtained shortly before and after the gamma-ray flare. The VERITAS observations are combined to obtain nightly $(E>200$ GeV) flux values since no evidence for intra-night variability is observed. The highest flux occurred on MJD 54749 and significant variability is observed during the whole interval ($\chi^{2}$ probability less than $10^{-4}$ for a fit of a constant flux).

The temporal dependence of the  {\em Fermi}-LAT  photon index and integral flux above 100 MeV and 1 GeV  are shown with time bins with width of 3 days in Figure \ref{longterm}. For those time intervals with no significant detection a 95\%-confidence flux upper limit is calculated. The flux  and  photon index  from the {\em Fermi}-LAT first source catalog  \citep{1FGL} are shown as horizontal lines for comparison. These values correspond to the average flux and photon index measured during the first eleven months of {\em Fermi} operations, and thus span the time interval considered in the figures. It is evident from the plot that the VHE flare detected by VERITAS starting on MJD 54740 is coincident with a period of  high flux in the {\em Fermi} energy band. The photon index during this time interval is consistent within errors with the average photon index   $\Gamma=1.95\pm0.03$ measured during the first 6 months of the {\em Fermi} mission \citep{LBASspectra}.

Long-term and  well-sampled light curves are available at optical and near-infrared wavelengths thanks to  observations by GASP-WEBT, ATOM, MDM and PAIRITEL.  Unfortunately, radio observations were too limited to obtain a light curve and no statement about variability in this band can be made. The best sampling is available for the {\em R} band, for which  variations  with a factor of $\gtrsim$ 2 are observed in the long-term light curve. Furthermore,  variability on time scales of less than a day is observed, as indicated in Figure \ref{shortterm}, and as previously reported by \citet{Webt_2007} following the  WEBT (Whole Earth Blazar Telescope) campaign on  3C~66A in 2007-2008.

\begin{figure}[htbp]
\begin{center}
\includegraphics[width=0.8\textwidth]{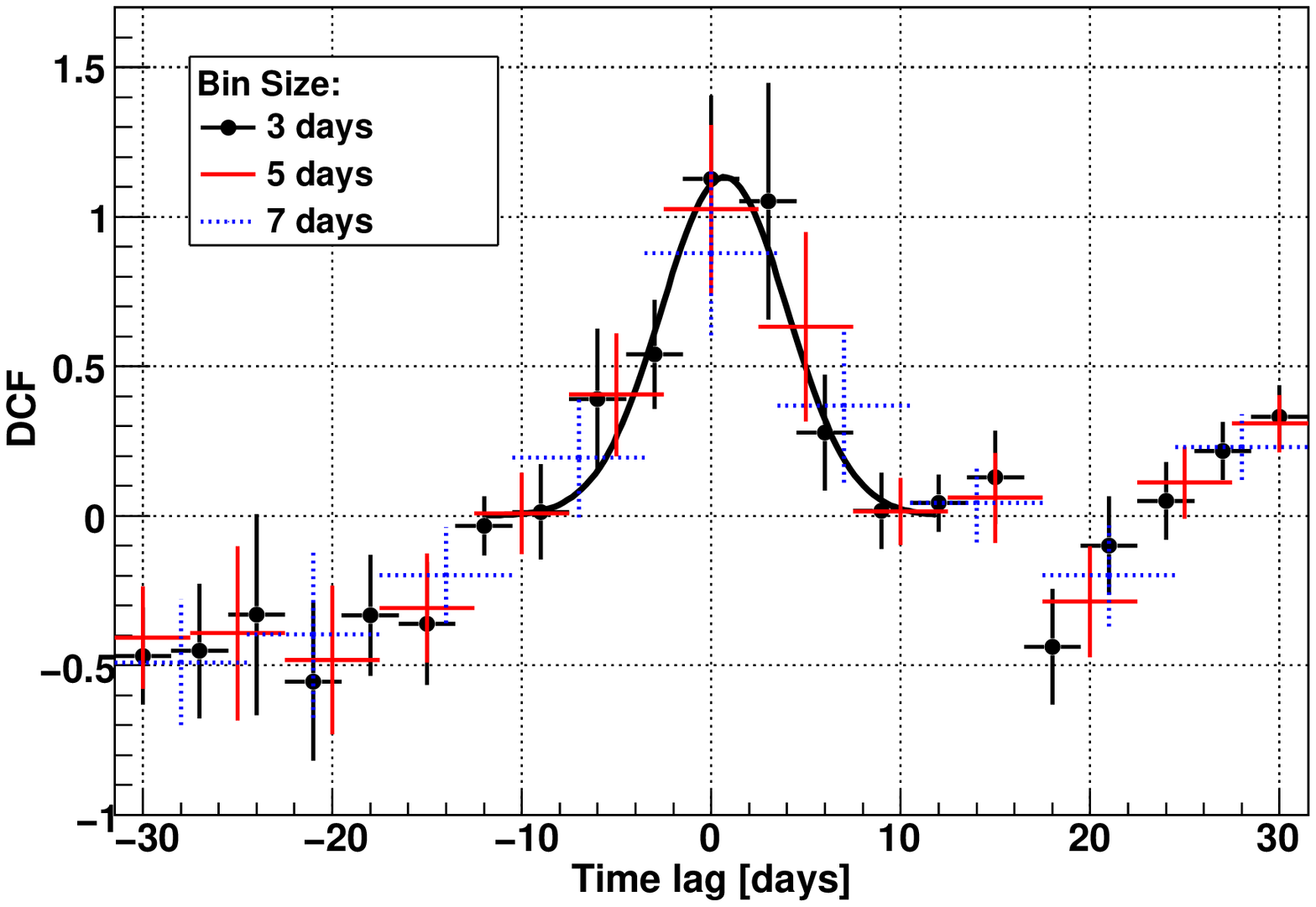}
\caption{{\footnotesize  Discrete correlation function (DCF) of the F(E $>$ 1 GeV) gamma-ray light curve with respect to the {\em R} band light curve. A positive time lag indicates that the gamma-ray band leads the optical band. Different symbols correspond to different bin sizes of time lag as indicated in the legend. The profile of the DCF is independent of bin size and is well described by a Gaussian function  of the form $DCF(\tau) = C_{max}\times \exp{(\tau - \tau_{0})^{2}/\sigma^{2}}$. The fit to the 3-day bin size distribution is shown in the plot as solid black line and the best-fit parameters are  $C_{max}=1.1\pm0.3$, $\tau_{0}=(0.7\pm0.7)$ days and $\sigma=(3.3\pm0.7)$ days. }}
\label{dcf}
\end{center}
\end{figure}

The increase in gamma-ray flux observed in the {\em Fermi} band  seems contemporaneous with a period of increased flux in the optical, and to test this hypothesis, the discrete correlation function (DCF)  is used \citep{dcf}. Figure \ref{dcf} shows the DCF of the F(E $>$ 1 GeV) gamma-ray band with respect to the {\em R} band with time-lag bins  of 3, 5 and 7 days. The profile of the DCF is consistent for all time-lag bins, indicating that the result is independent of bin size.  The DCF with time-lag bins of 3 days was fitted with a Gaussian function of the form $DCF(\tau) = C_{max}\times \exp{(\tau - \tau_{0})^{2}/\sigma^{2}}$, where $C_{max}$ is the peak value of the DCF,  $\tau_{0}$ is the delay timescale at which the DCF peaks, and $\sigma$ parametrizes the Gaussian width of the DCF.  The best fit function is plotted in Figure \ref{dcf} and the best fit parameters are  $C_{max}=1.1\pm0.3$, $\tau_{0}=(0.7\pm0.7)$ days and $\sigma=(3.3\pm0.7)$ days.  An identical analysis was also performed between the  F(E $>$ 100 MeV) and the {\em R} optical band with consistent results. This indicates a clear correlation between the {\em Fermi}-LAT and optical energy bands with a time lag that is consistent with zero and not greater than $\sim$5 days.  Despite the sparsity of the VERITAS light curve (due in part to the time periods when the source was not observable due to the full Moon) the DCF analysis was also performed  to search for correlations with either the {\em Fermi}-LAT or optical data. Apart from the overall increase in flux, no significant correlations can be established. {The onset of the E $>$ 200 GeV flare seems delayed by about $\sim$5 days with respect to the optical-GeV flare but given the coverage gaps no firm conclusion can be drawn (e.g., the flare could have been already underway when the observations took place). No such lag is expected from the homogeneous model described in the next section but could arise  in models with complex energy stratification and geometry in the emitting region.

\subsection{SED and Modeling}
  
The broadband SED derived from these observations is presented in Figure \ref{broadband} and modeled using the code of \citet{Boettcher_model}. In this
model, a power-law distribution of ultrarelativistic electrons
and/or pairs with lower and upper energy cutoffs at $\gamma_{\rm min}$
and $\gamma_{\rm max}$, respectively, and power-law index $q$
is injected into a spherical region of co-moving radius $R_B$. 
The injection rate is normalized to an injection luminosity
$L_e$, which is a free input parameter of the model. The model assumes  a 
temporary equilibrium between particle injection, radiative
cooling due to synchrotron and Compton losses, and particle
escape on a time  $t_{\rm esc} \equiv \eta_{\rm esc}
\, R_B/c$,  where $\eta_{\rm esc}$ is a scale parameter in the range  $\sim$ 250-500. Both the internal synchrotron photon
field (SSC) and external photon 
sources (EC) are considered as targets 
for Compton scattering. The emission region is moving with
a bulk Lorentz factor $\Gamma$ along the jet. To reduce the
number of free parameters, we assume that the jet is oriented
with respect to the line of sight at the superluminal angle
so that the Doppler factor is equal to $D = \left( \Gamma \, [1 - \beta
\, \cos\theta_{\rm obs} ] \right)^{-1}=\Gamma$, where $\theta_{\rm obs}$ is the angle of the jet with respect to the line of sight.  Given the uncertainty in the redshift 
determination of 3C~66A, a range of plausible redshifts, namely $z = 0.1, 0.2, 0.3,$ and the generally used catalog value $z = 0.444$ are considered for the modeling. All model fits  include  EBL absorption using the  optical depth values from \citet{franceschini08}.

\begin{figure}[htbp]
\begin{center}
\includegraphics[width=1.0\textwidth]{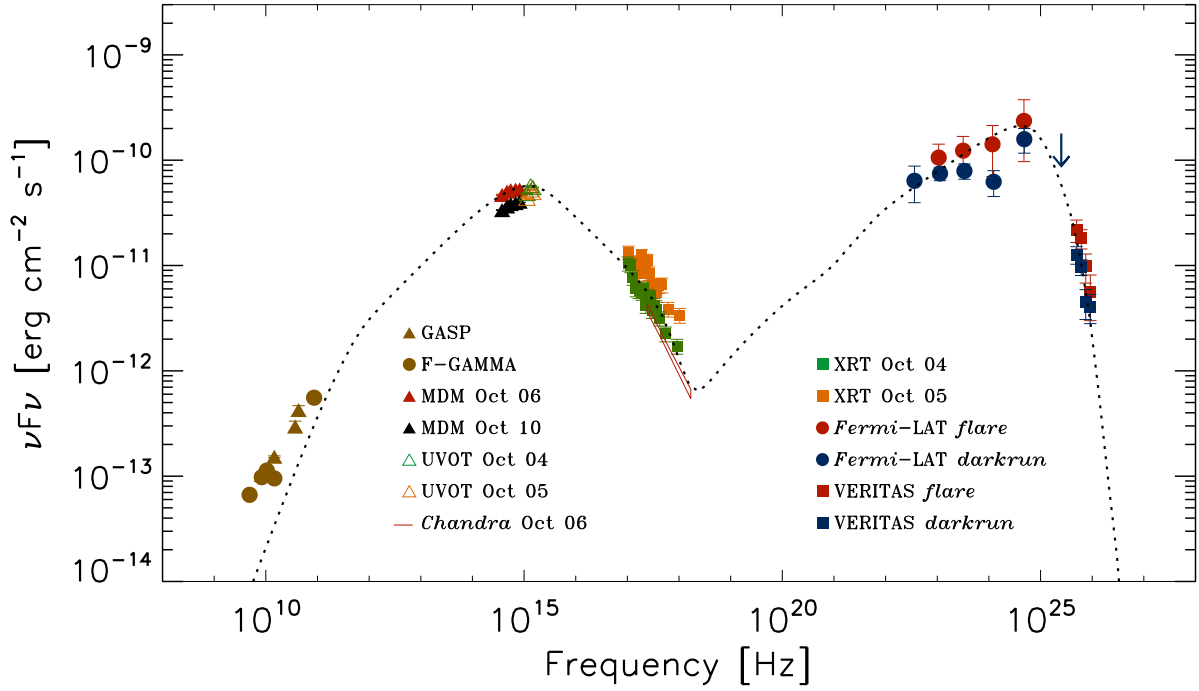}
\caption{{\footnotesize  Broadband SED of 3C~66A during the October 2008 multiwavelength campaign. The observation that corresponds to each set of data points is indicated in the legend. As an example, the EBL-absorbed EC+SSC model for $z=0.3$ is plotted here for reference. A description of the model is provided in the text.}}
\label{broadband}
\end{center}
\end{figure}

Most VHE blazars known to date are high synchrotron peaked blazars (HSPs), whose SEDs can often be fit satisfactorily with pure SSC models. Since the   transition from HSP to ISP is continuous,  a pure SSC model was fit first to  the radio through VHE gamma-ray SED.  Independently of the model under consideration, the low-frequency  part of the SED ($< 10^{20}$ Hz) is well fit with a synchrotron component, as shown in Figure \ref{broadband}. For clarity, only the high-frequency range is shown in Figures \ref{sed_ssc} and \ref{sed_ec}, where the different models are compared.   As can be seen from the figure,  a reasonable agreement with the overall SED can be achieved for any redshift in the  explored range. The weighted sum of squared residuals  has been calculated for the {\em Fermi}-LAT and VERITAS {\em flare} data (8 data points in total) in order to quantify the scatter of the points with respect to the model and is shown in Table \ref{SSCparameters}. The best agreement is achieved when the source is located at $z \sim0.2-0.3$. For lower redshifts, the 
model spectrum is systematically too hard, while at $z = 0.444$,  the 
model spectrum is invariably too soft as a result of  EBL absorption. It should be noted that the EBL model of \citet{franceschini08} predicts some of the lowest optical depth values in comparison to  other models \citep{Finke09_model, Gilmore09, Stecker06}. Thus, a model spectrum with redshift of 0.3 or above would be even harder to reconcile with the observations when using other EBL models.  

A major problem of the SSC models with $z\gtrsim0.1$ is that  $R_B$ is of the order of $\gtrsim$~5$\times10^{16}$~cm. This does not allow for variability
time scales shorter than $\lesssim$1 day, which seems to be in contrast with the
optical variability observed on shorter time scales.  A smaller $R_B$  would require an increase in the electron energy density (with no change in the magnetic field in order to preserve the flux level of the synchrotron peak) and would lead to internal gamma-gamma absorption. This problem could be mitigated by choosing extremely high 
Doppler factors, $D \gtrsim 100$. However, these are significantly larger than the values  inferred from VLBI observations of {\em Fermi}-LAT blazars \citep{Savolainen}\footnote{As a caveat, jet models with a decelerating flow \citep{GeorganopoulosKazanas2003, Piner2008} or with inhomogeneous transverse structure \citep{GhiselliniTavecchio2005, HenriSauge} can accommodate very high Doppler factors in the gamma-ray emitting region and still be consistent with the VLBI observations of the large scale jet.}. Moreover, all SSC models require very low
magnetic fields, far below the value expected from equipartition ($\epsilon_B=L_{B}/L_{e}\sim10^{-3}<<1)$, where $L_{B}$ is the Poynting flux derived from the magnetic energy density and $L_{e}$ is the energy flux of the electrons propagating along the jet). Table \ref{SSCparameters} 
lists the  parameters used for the SSC models displayed in Figure \ref{sed_ssc}.

\begin{figure}[htbp]
\begin{center}
\includegraphics[width=0.8\textwidth]{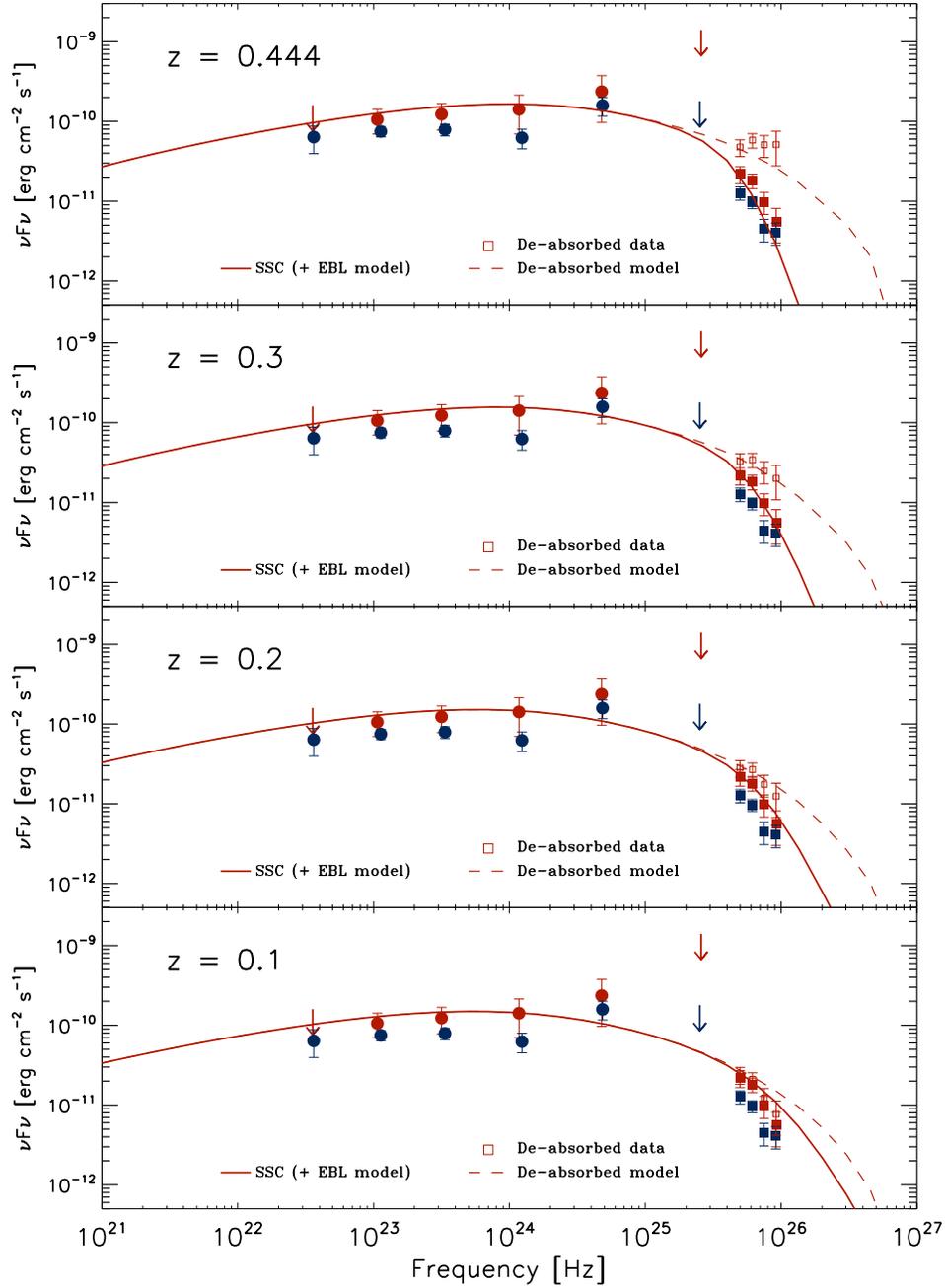}
\caption{{\footnotesize  SSC models for redshifts $z=$0.444, 0.3, 0.2, and 0.1 from top to bottom. The {\em Fermi}-LAT and VERITAS data points follow the same convention used in Figures \ref{gamma_sed} and \ref{broadband} to distinguish between {\em flare} (red) and {\em dark run} (blue) data points. In each panel the EBL-absorbed model  is shown as a solid red line and the de-absorbed model as a red dashed line. De-absorbed VERITAS {\em flare} points are shown as open squares. In all cases the optical depth values from \citet{franceschini08} are used.   The best agreement between the model and the data is  achieved when the source is located at $z = 0.2 - 0.3$.  For lower redshifts the model spectrum is systematically too hard, while at $z = 0.444$,  the model spectrum is too soft.}}
\label{sed_ssc}
\end{center}
\end{figure}

\begin{figure}[htbp]
\begin{center}
\includegraphics[width=0.8\textwidth]{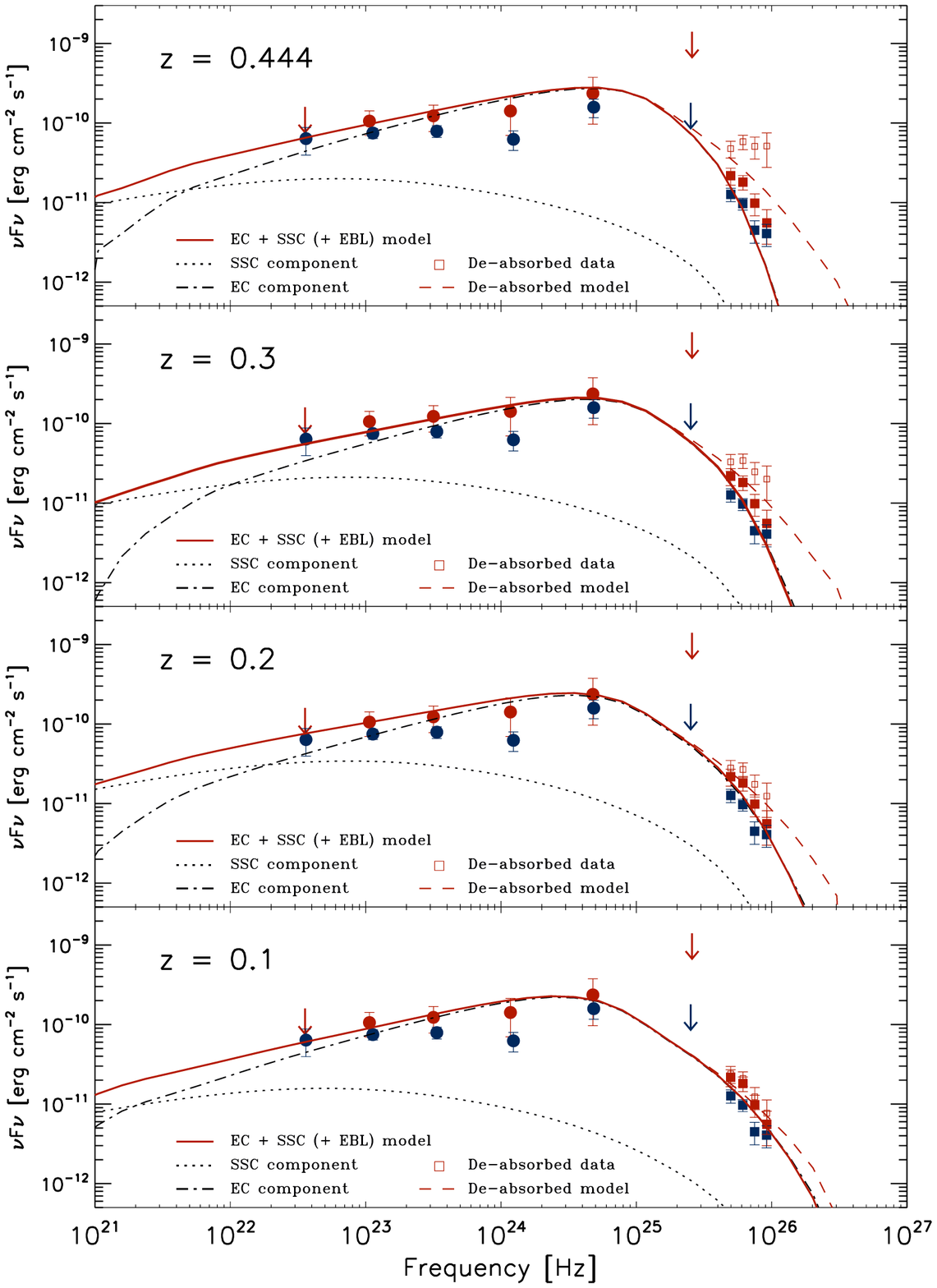}
\caption{{\footnotesize   EC + SSC model  for redshifts $z=$0.444, 0.3, 0.2, and 0.1 from top to bottom.  The individual EBL-absorbed EC and SSC components are indicated as a dash-dotted and dotted lines, respectively.  The sum is shown as solid red line (dashed when de-absorbed). The best agreement between the model and the data is  achieved when the source is located at $z \sim 0.2$.}}
\label{sed_ec}
\end{center}
\end{figure}

Subsequently, an external infrared radiation field with ad-hoc properties was included as a source of photons to be  Compton
scattered. For all SSC + EC models shown in Figure \ref{sed_ec},  the peak frequency of 
the external radiation field is set to $\nu_{\rm ext} = 1.4 \times 10^{14}$~Hz, corresponding 
to near-IR. This adopted value is high enough to produce E $\gtrsim$ 100 GeV photons from inverse Compton scattering
off the synchrotron electrons and at the same time is below the energy regime in which Klein-Nishina effects take place. 
Although the weighted sum of squared residuals  for EC+SSC models are generally worse than for pure SSC models, reasonable agreement with the overall SED can still be achieved for redshifts $z \lesssim 0.3$. Furthermore,  all  SSC + EC models are consistent with a variability time
scale of $\Delta t_{\rm var} \sim 4$~hr. This is in better agreement with the
observed variability at optical wavelengths than the pure SSC interpretation. Also, while the SSC + EC interpretation
still requires sub-equipartition magnetic fields, the magnetic fields are significantly
closer to equipartition than in the pure SSC case, with $L_{B}/L_{e}\sim 0.1$.  The parameters of the SSC + EC models are listed in Table \ref{ECparameters}.

Models with and without EC component yield the best agreement with the SED if the source is located at a redshift $z \sim 0.2$ -- 0.3. Of course, this depends on the EBL model used in the analysis. An EBL model that predicts higher attenuation than \citet{franceschini08} would lead to a lower  redshift range and make it even more difficult to have agreement between the SED models and the data when the source is located at redshifts $z\gtrsim0.4$.  Finally, it is worth mentioning that the redshift range $z \sim 0.2$ -- 0.3  is in agreement with previous estimates by \citet{finke2008}, who estimate the redshift of 3C~66A to be $z=0.321$ based on the magnitude of the host galaxy, and by  \citet{prandini} who use an empirical relation between the previously reported  {\em Fermi}-LAT and IACTs spectral slopes of blazars and their redshifts to estimate the redshift of 3C~66A to be  below $z=0.34\pm0.05$.

A detailed study of hadronic versus leptonic modeling of the October 2008 data will be published elsewhere, but it is worth mentioning that  the synchrotron proton blazar (SPB) model has been used to adequately reproduce the quasi-simultaneous SED observed during the 2003-2004 multi-wavelength campaign \citep{Reimer2008}. On that occasion rapid intraday variations down to ~2 hours time scale were observed, while during the 2008 campaign presented here these variations seem  less rapid. Qualitatively, the longer time scale variations may be due to a lower Doppler beaming, at the same time that a strongly reprocessed proton synchrotron component dominates the high energy output of this source. 

\section{Summary}

Multiwavelength observations of 3C~66A were carried out prompted by the gamma-ray outburst detected by the VERITAS and {\em Fermi} observatories in October 2008. This marks the first occasion that a gamma-ray flare is detected by GeV and TeV instruments in comparable time scales.  The light curves obtained  show strong variability at every observed wavelength and in particular, the flux increase observed by VERITAS and {\em Fermi} is coincident with an optical outburst.   The clear correlation between the {\em Fermi}-LAT  and  {\em R} optical light curves permits one to go beyond the source association reported in the 1st {\em Fermi}-LAT source  Catalog \citep{1FGL} and finally identify the gamma-ray source 1FGL J0222.6+4302 as blazar 3C~66A.     

For the modeling of the  overall SED a  reasonable agreement can be achieved using both a pure SSC model and an SSC + EC model with
an external near-infrared radiation field as an additional source for Compton scattering.
However, the pure SSC model requires (a) a large emission region, which is inconsistent
with the observed intra-night scale variability at optical wavelengths and (b) low magnetic fields, about a factor
$\sim 10^{-3}$ below equipartition. In contrast, an SSC + EC interpretation allows
for variability on time scales of a few hours, and for magnetic fields within about an
order of magnitude of, though still below, equipartition. It is worth noting that the results presented here agree with the findings following the $(E>200$ GeV) flare of blazar W Comae (also an ISP) in  2008 March \citep{Wcomae}. In both cases the high optical luminosity is expected to  play a key role in providing the seed population for IC scattering.

Intermediate synchrotron peaked blazars like 3C~66A are well suited for  simultaneous observations by {\em Fermi}-LAT and ground-based IACTs like VERITAS. Relative to the sensitivities of these instruments,  ISPs are bright enough to allow for time-resolved spectral measurements in each band during flaring episodes. These types of observations coupled with extensive multi-wavelength coverage at lower energies will continue to provide key tests of blazar emission models.  

\begin{table}[p]
\begin{center}\small
\begin{tabular}{|c|c|c|c|c|} 
\hline
Model parameter                          & $z=$ 0.1  & $z=$ 0.2  & $z=$ 0.3  & $z=$ 0.444   \\
\hline \hline
Low-energy cutoff $(\gamma_{\rm min})$         & 1.8$\times10^4$   & 2.0$\times10^4$   & 2.2$\times10^4$    & 2.5$\times10^4$    \\
High-energy cutoff $(\gamma_{\rm max})$         & 3.0$\times10^5$    & 4.0$\times10^5$  & 4.0$\times10^5$   & 5.0$\times10^5$   \\
Injection index $(q)$                                 & 2.9   & 2.9  & 3.0  & 3.0  \\
Injection luminosity $(L_e)$ [$10^{45}$ erg s$^{-1}$]      & 1.3  & 3.3 & 5.7  & 12.8  \\
Co-moving magnetic field $(B)$ [G]              & 0.03 & 0.02 & 0.02 & 0.01 \\
Poynting flux $(L_B)$ [$10^{42}$ erg s$^{-1}$]      & 1.1  & 4.9 & 8.5  & 13.7  \\
$\epsilon_B \equiv L_B/L_e$       & 0.9$\times10^{-3}$  & 1.5$\times10^{-3}$ & 1.5$\times10^{-3}$  & 1.1$\times10^{-3}$  \\
Doppler factor $(D)$                  & 30    & 30   & 40    & 50    \\
Plasmoid radius $(R_B)$ [$10^{16}$ cm]                    & 2.2   & 6.0  & 7.0   & 11    \\
Variability time scale $(\delta t_{\rm var}^{\rm min})$ [hr] & 7.4  & 22.1 & 21.1  & 29.4  \\
\hline
\hline
{\footnotesize Weighted sum of squared residuals  to VERITAS {\em flare} data}   & 7.1 & 0.9 & 0.7 & 6.2 \\
{\footnotesize Weighted sum of squared residuals  to {\em Fermi}-LAT {\em flare} data}  & 1.6 & 1.6 & 1.3 & 1.4 \\
Total weighted sum of squared residuals   & 8.7 & 2.5 & 1.9 & 7.6 \\
\hline
\end{tabular}\\
\end{center}
\caption{\label{SSCparameters}Parameters used for the SSC models displayed in Figure \ref{sed_ssc}. All SSC models require very low
magnetic fields, far below the value expected from equipartition (i.e. $\epsilon_B<<1$). The  weighted sum of squared residuals to the VERITAS and  {\em Fermi}-LAT data and the total value for the combined data set  are included at the bottom of the table. The best agreement between the model and the data is obtained when the source is at redshift $z= 0.2 - 0.3$.  }
\end{table}

\begin{table}[p]
\begin{center}\small
\begin{tabular}{|c|c|c|c|c|} 
\hline
Model parameter                          & $z=$ 0.1  & $z=$ 0.2  & $z=$ 0.3  & $z=$ 0.444   \\
\hline \hline
Low-energy cutoff  $(\gamma_{\rm min})$              & 5.5$\times10^3$   & 7.0$\times10^3$   & 6.5$\times10^3$   & 6.0$\times10^3$   \\
High-energy cutoff $(\gamma_{\rm max})$              & 1.2$\times10^5$   & 1.51.2$\times10^5$   & 1.51.2$\times10^5$   & 1.51.2$\times10^5$   \\
Injection index    $(q)$                               & 3.0   & 3.0   & 3.0  & 3.0   \\
Injection luminosity $(L_{e})$ [$10^{44}$ erg s$^{-1}$]          & 1.1  & 4.2  & 6.0  & 10.4  \\
Co-moving magnetic field $(B)$ [G]                  & 0.35  & 0.22  & 0.21  & 0.23  \\
Poynting flux $(L_B)$ [$10^{43}$ erg s$^{-1}$]          & 1.0  & 2.4  & 6.0  & 11.2  \\
$\epsilon_B \equiv L_B/L_e$             & 0.10 & 0.06 & 0.10  & 0.11  \\
Doppler factor    $(D)$                      & 30    & 30    & 40    & 50    \\
Plasmoid radius $(R_{B})$ [$10^{16}$ cm]                        & 0.5   & 1.2   & 1.5   & 1.5   \\
Variability time scale $(\delta t_{\rm var}^{\rm min})$ [hr]     & 1.7  & 4.4  & 4.5  & 4.0  \\
Ext. radiation energy density  [$10^{-6}$ erg cm$^{-3}$] & 5.4   & 2.4   & 1.2   & 1.3   \\
\hline
\hline
{\footnotesize Weighted sum of squared residuals  to VERITAS {\em flare} data}   & 4.8 & 3.6 & 7.9 & 15.7 \\
{\footnotesize Weighted sum of squared residuals  to {\em Fermi}-LAT {\em flare} data}   & 1.0 & 1.2 & 0.8 & 1.5 \\
Total weighted sum of squared residuals   & 5.8 & 4.8 & 8.7 & 17.2 \\
\hline
\end{tabular}\\
\end{center}
\caption{\label{ECparameters}Parameters used for the EC+SSC model fits displayed in Figure \ref{sed_ec}.  These model fits require magnetic fields closer to equipartition and allow for the intra-night variability observed in the optical data. The weighted sum of squared residuals to the VERITAS and  {\em Fermi}-LAT data and the total value for the combined data set are included at the bottom of the table. }
\end{table}

\acknowledgments

\section*{Acknowledgements}

The \textit{Fermi} LAT Collaboration acknowledges generous ongoing support
from a number of agencies and institutes that have supported both the
development and the operation of the LAT as well as scientific data analysis.
These include the National Aeronautics and Space Administration and the
Department of Energy in the United States, the Commissariat \`a l'Energie Atomique
and the Centre National de la Recherche Scientifique / Institut National de Physique
Nucl\'eaire et de Physique des Particules in France, the Agenzia Spaziale Italiana
and the Istituto Nazionale di Fisica Nucleare in Italy, the Ministry of Education,
Culture, Sports, Science and Technology (MEXT), High Energy Accelerator Research
Organization (KEK) and Japan Aerospace Exploration Agency (JAXA) in Japan, and
the K.~A.~Wallenberg Foundation, the Swedish Research Council and the
Swedish National Space Board in Sweden. Additional support for science analysis during the operations phase is gratefully
acknowledged from the Istituto Nazionale di Astrofisica in Italy and the Centre National d'\'Etudes Spatiales in France.  

The VERITAS collaboration acknowledges the generous support from the US Department of Energy, the US National Science Foundation, and the Smithsonian Institution, by NSERC in Canada, by Science Foundation Ireland, and by STFC in the UK. The VERITAS collaboration also acknowledges the excellent work of the technical support staff at the FLWO and the collaborating institutions in the construction and operation of the instrument, as well as support from the NASA/{\em Swift} guest investigator program (grant NNX08AU13G) for the {\em Swift} observations.

PAIRITEL is operated by the Smithsonian Astrophysical Observatory (SAO) and was made possible by a grant from the Harvard University Milton Fund, a camera loan from the University of Virginia, and continued support of the SAO and UC Berkeley. The PAIRITEL project is further supported by NASA/{\em Swift} Guest Investigator grant NNG06GH50G. This research is partly based on observations with the 100-m telescope of the MPIfR (Max-Planck-Institut f\"ur Radioastronomie) at Effelsberg and has also made use of observations with the IRAM 30-m telescope. The Mets\"ahovi team acknowledges the support from the Academy of Finland. The Abastumani Observatory team acknowledges financial support by the
Georgian National Science Foundation through grant GNSF/ST08/4-404. The St. Petersburg University team acknowledges support from Russian RFBR foundation via grant 09-02-00092.
AZT-24 observations are made within an agreement between  Pulkovo, Rome and Teramo observatories.

L.~C. Reyes acknowledges the support by the Kavli Institute for Cosmological Physics at the University of Chicago through grants NSF PHY-0114422 and NSF PHY-0551142 and an endowment from the Kavli Foundation and its founder Fred Kavli. M. B{\"o}ttcher acknowledges support from NASA through Chandra Guest Investigator Grant GO8-9100X. Some of the VERITAS simulations used in this work have been performed on the joint Fermilab - KICP supercomputing cluster, supported by grants from Fermilab, the Kavli Institute for Cosmological Physics, and the University of Chicago.

\clearpage

\end{document}